\documentclass[prl,twocolumn,showpacs]{revtex4}
\usepackage{graphicx}
\usepackage{psfig}
\begin{document}
\def\<{\langle}
\def\>{\rangle}
\def\tr{\mbox{Tr}}
\def\i{{\rm i}}
\def\e{{\rm e}}
\def\G{{\cal G}}
\def\L{{\cal L}}
\def\f{{\rm f}}
\title{Shot noise from action correlations} 
\author{Holger Schanz}\email{holger@chaos.gwdg.de}
\author{Mathias Puhlmann}
\author{Theo Geisel}
\affiliation{Max-Planck-Institut f\"ur Str\"omungsforschung und Institut
f{\"u}r Nichtlineare Dynamik der Universit{\"a}t G{\"o}ttingen, 
Bunsenstra{\ss}e 10, D-37073 G\"ottingen, Germany}
\date{\today}
\begin{abstract}
  We consider universal shot noise in ballistic chaotic cavities from a
  semiclassical point of view and show that it is due to action
  correlations within certain groups of classical trajectories. Using quantum
  graphs as a model system we sum these trajectories analytically and find
  agreement with random-matrix theory.  Unlike all action correlations which
  have been considered before, the correlations relevant for shot noise
  involve four trajectories and do not depend on the presence of any symmetry.
\end{abstract}
\maketitle

One of the most prominent methods to describe spectral and transport
properties of ballistic quantum systems with classically chaotic dynamics
relies on the semiclassical representation of the Green's function in terms of
classical trajectories. For closed systems this leads to Gutzwiller's trace
formula \cite{Gut90}, which expresses the oscillating density of states as sum
over periodic orbits. For open systems the semiclassical theory of chaotic
scattering \cite{Mil74,BS88}, and in particular its applications to electronic
transport through mesoscopic devices \cite{JBS90}, are based on this approach.
The power of this method as compared, e.~g., to random-matrix theory (RMT)
\cite{Bee97,GMW98} is its potential to account for system-specific details.
Its main drawback lies in the difficulty to handle the resulting sums over
huge sets of classical trajectories. Until recently the only method to deal
with this problem was Berry's dia\-go\-nal approximation \cite{Ber85}, which
neglects any nontrivial correlations between the trajectories. As a
consequence many interesting phenomena such as weak localization, universality
in spectral statistics and in conductance fluctuations, or the supression of
shot noise cannot be described properly.  Although the role of correlations
between the actions of classical orbits has been appreciated for a long time
[8--10], there is only one special case where they can be
accounted for explicitly: Sieber and Richter recently calculated the leading
order weak localization corrections from correlations between specific orbit
pairs \cite{SR01}. This result stimulated further intense research
\cite{BSW02,BHH02} and it is clearly a very promising approach. However, as
mentioned above, weak localization is just one besides a variety of other
phenomena for which action correlations are relevant as well, but cannot be
accounted for by the orbit pairs considered in
[11--13].

In this paper we address shot noise in ballistic mesoscopic conductors, which
is an important source of (experimentally accessible) information about the
dynamics in such systems 
[18--21]. We identify the action
correlations which are responsible for shot noise and explain how due to these
correlations the universal result of RMT
[5, 14--20] can be recovered from a
single system without ensemble average. The relevant correlations are
fundamentally different from all those considered previously
[8--13] because they involve four instead of
just two classical trajectories. Moreover they do not depend on the presence of
symmetries. We emphasize that action correlations are no small
correction. They are needed to understand shot noise {\em to leading order}.

\begin{figure}[htb]
  \centerline{\psfig{figure=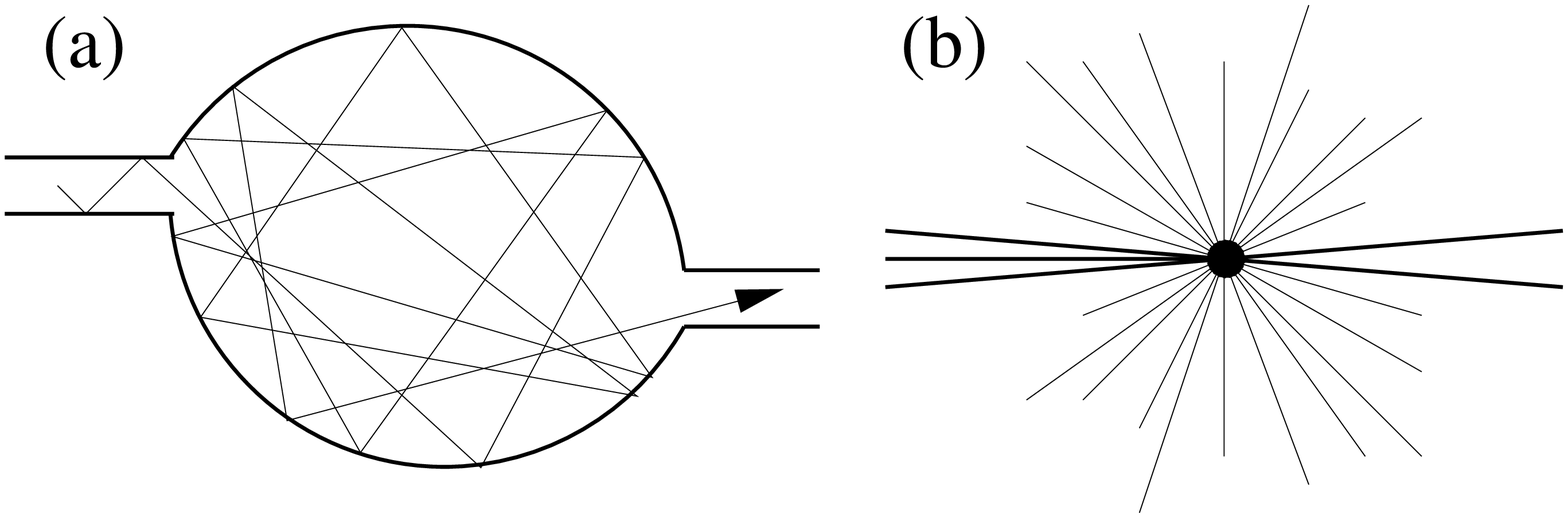,width=8cm}}
 \caption{\label{cavity} (a) A chaotic cavity with two attached waveguides and
   a classical trajectory contributing to conductance and shot noise.
   (b) The quantum graph used to model this situation. The transversal
   modes of the waveguides correspond to the infinite leads attached to the
   graph (bold lines), while the internal system is represented by many finite
   bonds.}
\end{figure}

We will perform all explicit calculations in a specific model system, the
quantum graph of Fig.~\ref{cavity}b. Quantum graphs (networks) have a long
record as models for electronic transport (see \cite{Kuc02} and
Refs.~therein). Since the pioneering work of Kottos and Smilansky they are
also established in quantum chaos 
[23--25, 12].  They are
particularly suitable for our purpose as the representation in terms of
classical trajectories is exact and also the analogue of action correlations
amounts to exact degeneracies. Previous studies showed that despite these
analytical simplifications the mechanism and the role of correlations between
classical trajectories are equivalent to other systems such as billiards
\cite{BSW02}.

Consider a chaotic cavity with two attached wave\-guides supporting $N_{1}$,
$N_{2}$ transversal modes, respectively (Fig.~\ref{cavity}a). For small bias
voltage and temperature, negligible electron interactions and fully coherent
dynamics all information about electron transport through this system is
contained in the scattering matrix at the Fermi energy
\begin{equation}\label{s}
{\cal S}=\pmatrix{r&t'\cr t&r'}\,.
\end{equation}
Shot noise represents temporal current fluctuations due to the discreteness of
the electron charge \cite{Sch18}. At zero temperature the average power of the
noise can be expressed in terms of the transmission matrix $t$ as
\cite{But90,Bee97}
\begin{equation}\label{P}
P=2e|V|\,G_{0}\;\tr\,tt^{\dagger}(1-tt^{\dagger})\,,
\end{equation}
while the assumption of uncorrelated electrons yields $P_{\rm
  Poisson}=2e|V|\,G$. Here $G=G_0\;\tr\,tt^{\dagger}$ denotes the
conductance, $e$ is the electron charge, $V$ the voltage and $G_{0}=2e^2/h$
the conductance quantum. RMT yields \cite{Bee97}
\begin{equation}\label{RMT}
\<P\>_{\rm RMT}=2e|V|\,G_{0}\; {N_{1}^2N_{2}^{2}/ (N_{1}+N_{2})^{3}}
\end{equation}
and for the conductance $\<G\>_{\rm RMT}=G_{0}\,N_{1}N_{2}/(N_{1}+N_{2})$.
Each result comes with a weak localization correction which is small
($\<\delta P\>\ll \<P\>$, $\<\delta G\>\ll \<G\>$) for large
$N_{1}$, $N_{2}$ and will not be considered here. For $N_{1}=N_{2}$ the
Fano factor $F=\<P\>/\<P_{\rm Poisson}\>=1/4$ is obtained.

\begin{figure}[htb]
  \centerline{\psfig{figure=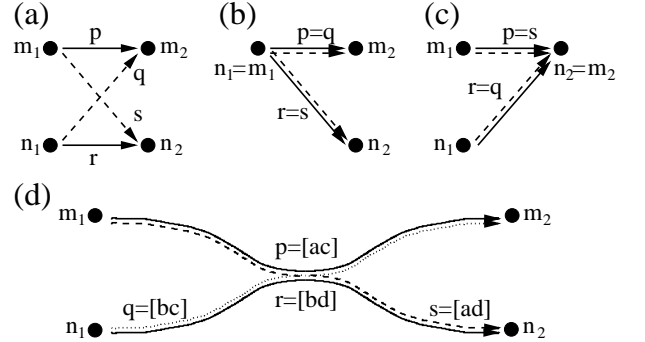,width=8cm}}
 \caption{\label{arrows} (a) shows schematically four classical trajectories
   $p,q,r,s$ connecting two incoming modes $m_{1},n_{1}$ to two outgoing modes
   $m_{2},n_{2}$ such that the diagram contributes to the semiclassical
   approximation of Eq.~(\protect\ref{P}). A contribution to Eq.~(\ref{t4})
   results only if the trajectories are correlated such that the action
   difference between $p,r$ (full lines) and $q,s$ (dashed lines) remains
   small for variying energy. (b) and (c) shows the simplest configurations
   where this is the case: two trajectories are pairwise equal (diagonal
   approximation). (d) shows the configuration which completely accounts for
   the universal shot noise in leading order.}
\end{figure}

We will show that Eq.~(\ref{RMT}) can be recovered semiclassically.  To this
end, the element $t_{n_{2}n_{1}}$ of the transmission matrix is expressed as a
sum over all classical trajectories\footnote{We emphasize that these
  trajectories are {\em deterministic}. Therefore our approach must not be
  confused with other quasiclassical theories such as
  \protect\cite{BS00,NSP02,AAL00}.} connecting the incoming mode $1\le
n_{1}\le N_{1}$ with the outgoing mode $1\le n_{2}\le N_{2}$ \cite{BJS93}
\begin{equation}\label{t}
t_{n_{2}n_{1}}=\sum_{p}A_{p}\,\e^{\i S_{p}/\hbar}\,,
\end{equation}
where $S_{p}$ denotes the classical action and $A_{p}$ is an amplitude related
to the stability of the trajectory. While for the conductance we have to
evaluate a sum over pairs of trajectories
\begin{equation}\label{t2}
\<\tr\,tt^{\dagger}\>=
\sum_{n_{1}=1}^{N_{1}}\sum_{n_{2}=1}^{N_{2}}
\sum_{pq:n_{1}\to n_{2}}A_{p}A_{q}^{*}\,
\<\e^{\i/\hbar(S_{p}-S_{q})}\>\,,
\end{equation}
the shot noise involves also a term combining four classical paths
\begin{eqnarray}\label{t4}
\<{\rm Tr}\,(tt^{\dagger})^{2}\>&=&
\sum_{m_{1}n_{1}}\sum_{m_{2}n_{2}}
\sum_{pqrs}A_{p}A_{q}^{*}A_{r}A_{s}^{*}
\nonumber\\&&\qquad\qquad\times
\<\e^{\i/\hbar(S_{p}-S_{q}+S_{r}-S_{s})}\>\,.
\end{eqnarray}
Here, the trajectories $p,q,r,s$ connect two incoming to two outgoing modes as
shown in Fig.~\ref{arrows}a. As Eqs.~(\ref{t2}), (\ref{t4}) describe one
particular system rather than an ensemble, the average $\<\cdot\>$ is to be
taken over an energy window. It should be small enough to keep the classical
dynamics and in particular the amplitudes $A_{p}$ essentially unchanged.
Nevertheless, in the semiclassical limit $\hbar\to 0$ the phase factor is
rapidly oscillating and only those orbit combinations survive the averaging
for which the action changes are correlated.

In particular, setting $p=q$ in Eq.~(\ref{t2}) the phase drops out 
and we are left with a sum over classical probabilities $|A_{p}|^{2}$. This is
the diagonal approximation \cite{Ber85}.  Provided that the dwell time inside
the open cavity is large compared to the time needed for equidistribution over
the available phase space, the probability is the same for all outgoing modes
and $\<G\>_{\rm RMT}$ is exactly recovered \cite{JBS90}. Hence, the
contribution from other pairs of correlated trajectories that might exist must
vanish although no explicit demonstration of this fact has been given up to
now. In presence of time-reversal symmetry the above remains valid to leading
order in the mode number $N$.

For shot noise the diagonal approximation has two analogues
(Fig.~\ref{arrows}b,c): we can have (i) $p=q$, $r=s$ for $m_{1}=n_{1}$ and
(ii) $p=s$, $r=q$ for $m_{2}=n_{2}$. In both cases no phases are left in
Eq.~(\ref{t4}) and the remaining summation is over two independent classical
trajectories $p$, $r$ with the only constraint that they begin or end at the
same mode, respectively.  Proceeding as in the diagonal
approximation to the conductance we obtain \footnote{ Note that terms with
  $p=q=r=s$ are contained in Eq.~(\ref{t4di}) but also in Eq.~(\ref{t4dii}).
  To avoid this overcounting one should, e.~g., restrict the second sum to
  $n_{1}\ne m_{1}$ but the resulting lower-order correction is not consider
  here.}
\begin{equation}\label{t4di}
\sum_{m_{1}=1}^{N_{1}}\sum_{m_{2},n_{2}=1}^{N_{2}}
\sum_{pr}|A_{p}|^{2}\,|A_{r}|^{2}={N_{1}N_{2}^{2}\over (N_{1}+N_{2})^{2}}\,,
\end{equation}
\begin{equation}\label{t4dii}
\sum_{m_{1},n_{1}=1}^{N_{1}}\sum_{m_{2}=1}^{N_{2}}
\sum_{pr}|A_{p}|^{2}\,|A_{r}|^{2}={N_{1}^{2}N_{2}\over (N_{1}+N_{2})^{2}}\,.
\end{equation}
Combining these two results we find 
\begin{equation}\label{noshot}
\<\tr\,(tt^{\dagger})^{2}\>_{\rm diag}=
\<\tr\,tt^{\dagger}\>_{\rm diag}=
N_{1}N_{2}/(N_{1}+N_{2})
\end{equation}
and according to Eq.~(\ref{P}) this means that within the diagonal
approximation there is no shot noise. This is no surprise: the diagonal
approximation reduces the quantum to the classical problem and since classical
dynamics is deterministic there is no uncertainty if an incoming electron is
transmitted or not and hence no noise \cite{BV91,AAL00}. On the other hand
Eq.~(\ref{noshot}) is quite remarkable as it means that within the
semiclassical approximation {\em shot noise is entirely due to nontrivial
  correlations between different trajectories}.

What is the general mechanism for such correlations?  
Previous research 
[9--13] showed that pairs of
trajectories have correlated actions if they explore the same (or
symmetry-related) parts of phase space with a different
itinerary. In terms of symbolic dynamics the code words
for the two orbits are composed of the same sequences, in permuted order. The
analogy to diagrammatic perturbation theory and some recent results
\cite{SR01,BSW02} suggest further that the importance of the correlations
decreases with growing number of sequences needed to represent the code of the
trajectories: in the diagonal approximation to the conductance the codes are
equal and the result is correct to leading order, orbit pairs composed of two
loops give the next-to-leading order correction etc.

In the case of shot noise we have exhausted the diagonal approximation and
consider therefore trajectories $p,q,r,s$ whose codes can all be represented
in terms of two subsequences. Inspection shows that the only option is the
diagram of Fig.~\ref{arrows}d.  Indeed the phase in Eq.~(\ref{t4}) will almost
vanish for such contributions since the combination of $p,r$ (full lines)
almost coincides with the combination of $q,s$ (dashed lines) such that the
respective actions cancel. A remaining small total action difference comes
from the different behaviour of the trajectories inside the crossing
region. In this respect the correlated trajectories of Fig.~\ref{arrows}d are
very similar to those giving rise to weak localization effects
[11--13], i.~e.\ the methods developed there for various
specific systems should allow for a straightforward generalizion to shot
noise.

In the remainder of this paper we treat one of those systems explicitly,
namely the quantum graph shown in Fig.~\ref{cavity}b. The closed version of
this graph consists of a central vertex with valency $B$ and $b=1\dots B$
attached bonds with incommensurate lengths $L_{b}$. Following the standard
quantization \cite{KS97} the dynamics of a particle with wavenumber
$k=(2mE)^{1/2}/\hbar$ is represented by a $B\times B$ bond-scattering matrix
$\Sigma_{bb'}(k)=\sigma_{bb'}\e^{2\i kL_{b}}$ containing energy-dependent
phases $2kL_{b}$ from the free motion on the bonds and complex amplitudes
describing the scattering at the central vertex. A basic requirement is
unitarity (current conservation)
\begin{equation}\label{unitarity}
\delta_{\alpha\alpha'}=\sum_{\beta=1}^{B}
\sigma_{\beta\alpha}\sigma^{*}_{\beta\alpha'}\,,
\end{equation}
otherwise $\sigma$ can be chosen according to the physical situation.  For
simplicity we set 
\begin{equation}\label{sigma}
\sigma_{bb'}=\e^{2\pi\i bb'/B}/\sqrt{B}
\end{equation}
such that all classical transition probabilities are equal
$|\sigma_{bb'}|^2\equiv 1/B$.  This model was first considered by Tanner
\cite{Tan01} who showed numerically that its specral statistics follows RMT.
We open the graph by extending $N=N_{1}+N_{2}$ bonds to infinity and model a
two-channel geometry by considering $N_{1}$ ($N_{2}$) of these leads as the
modes in the left (right) contact (Fig.~\ref{cavity}b). For fixed $N_{1}$,
$N_{2}$ we will consider the limit $B\to\infty$ in order to meet the condition
of a long dwell time which was already mentioned in connection with the
diagonal approximation. Within the leading-order in $B$ we also neglect
lower-order corrections in the mode numbers $N_{1}$, $N_{2}$ in order to
compare our result to Eq.~(\ref{RMT}).

For a graph the $N\times N$ unitary scattering matrix Eq.~(\ref{s}) can be
expressed in terms of subblocks of the bond-scattering matrix $\Sigma$ via
${\cal S}=\Sigma_{\L\L}+\Sigma_{\L\G}(I-\Sigma_{\G\G})^{-1}\Sigma_{\G\L}$
\cite{KS97}.  $\L=\L_{1}\cup\L_{2}$ denotes here the set of
$N=N_{1}+N_{2}$ leads and $\G$ comprises the $B-N$ bonds inside the graph.
Expanding the Greens function of the internal part $(I-\Sigma_{\G\G})^{-1}$
into a geometric series we arrive at Eq.~(\ref{t}) which is in the case of
graphs an identity rather than a semiclassical approximation. The sum is over
all trajectories (=bond sequences) $p=[n_{1}p_{1}\dots p_{t}n_{2}]$ connecting
the lead $n_{1}\in \L_{1}$ to the lead $n_{2}\in \L_{2}$ via an arbitrary
number $t\ge 0$ of internal bonds $p_{j}\in \G$. The action is related to the
total length of the trajectory, $S_{p}/\hbar=kL_{p}$, where
$L_{p}=\sum_{j=1}^{t}L_{p_{j}}$.  Finally the amplitude is given as
$A_{p}=\sigma_{n_{1}p_{1}}\sigma_{p_{1}p_{2}}\dots\sigma_{p_{t}n_{2}}$ such
that the classical probability of the trajectory is $|A_{p}|^{2}=1/B^{t+1}$.
Consequently, we have
\begin{equation}\label{classsum}
\sum_{p}|A_{p}|^{2}=\sum_{t=0}^{\infty}{(B-N)^{t}\over B^{t+1}}=
{1\over N}\,,
\end{equation}
from which we do indeed recover the diagonal approximation
Eqs.~(\ref{t4di})-(\ref{noshot}).

Next we consider trajectories $p,q,r,s$ which are composed of four sequences
$a,b,c,d$ as shown in Fig.~\ref{arrows}d and assume that each of these
sequences has a length $t\ge 1$. In order to avoid overcounting we have to
ensure that for any given set $p,q,r,s$ the definition of $a,b,c,d$ is unique.
Potential problems arise if all four trajectories coincide in the crossing
region for one (or more) steps: $p=[a\gamma c]$, $q=[b\gamma c]$, $r=[b\gamma
d]$, $s=[a\gamma d]$. It is a matter of taste if $\gamma$ in this situations
is considered as part of $a$ and $b$ or of $c$ and $d$. We use the first
representation and enforce it by the restriction $c_{\i}\ne d_{\i}$ (The
subscripts i/f are used for the initial/final bond in $a,b,c,d$.).  

Returning to Eq.~(\ref{t4}) we note that the actions of $p,r$ and $q,s$ cancel
exactly such that the phase factor is absent \footnote{This is not specific
  for the correlated trajectories considered here. Any correlations in graphs
  amount to exact degeneracies because the energy intervall for averaging in
  Eqs.~(\protect\ref{t2}), (\ref{t4}) can be chosen arbitrarily large
  \protect\cite{KS97,SS00}.}.  As in Eq.~(\ref{classsum}) we can perform the
summation over all internal bonds of the subsequences $a,b,c,d$ and also over
the leads $m_{1},n_{1},m_{2},n_{2}$. Only the amplitudes from transitions
right at the intersection of $a,b,c,d$ do not combine into classical
probabilities and must be considered explicitly. We obtain
\begin{eqnarray}\label{essum}
&&
t^{(4)}_{abcd}
=
{N_{1}^{2}N_{2}^{2}\over N^4}
\sum_{a_{\f}b_{\f},c_{i}\ne d_{\i}\in\G}\hspace*{-2mm}
\sigma_{a_{\f}c_{\i}}\sigma_{b_{\f}c_{\i}}^{*}
\sigma_{b_{\f}d_{\i}}\sigma_{a_{\f}d_{\i}}^{*}
\\
&&
={N_{1}^{2}N_{2}^{2}\over N^4}
\sum_{a_{\f}b_{\f}\in\L}\sum_{c_{\i}d_{\i}\in\G}
\sigma_{a_{\f}c_{\i}}\sigma_{b_{\f}c_{\i}}^{*}
\sigma_{b_{\f}d_{\i}}\sigma_{a_{\f}d_{\i}}^{*}(1-\delta_{c_{\i}d_{\i}})\,.
\nonumber\\&&
={N_{1}^{2}N_{2}^{2}\over N^4}\Big(
\sum_{a_{\f}=b_{\f}\in\L}\sum_{c_{\i}d_{\i}\in\G}
\label{step2}\\&&
\phantom{=}+\sum_{a_{\f}\ne b_{\f},c_{\i}d_{\i}\in\L}-
\sum_{a_{\f}\ne b_{\f}\in\L}\sum_{c_{\i}=d_{\i}\in\G}\Big)
\sigma_{a_{\f}c_{\i}}\sigma_{b_{\f}c_{\i}}^{*}
\sigma_{b_{\f}d_{\i}}\sigma_{a_{\f}d_{\i}}^{*}
\nonumber\\&&\label{final}
={N_{1}^{2}N_{2}^{2}\over N^3}+{\cal O}(B^{-1})\,.
\end{eqnarray}
To perform this calculation we have repeatedly used Eq.~(\ref{unitarity})
in a form which allows to transfer summations from the graph $\G$ to the
leads $\L$   
\begin{equation}\label{trick}
\sum_{\beta\in \G}\sigma_{\beta\alpha}\sigma^{*}_{\beta\alpha'}
=\delta_{\alpha\alpha'}-
\sum_{\beta\in \L}\sigma_{\beta\alpha}\sigma^{*}_{\beta\alpha'}\,.
\end{equation}
Moreover we used that according to Eq.~(\ref{sigma}) we have
$|\sigma_{a_{\f}c_{\i}}\sigma_{b_{\f}c_{\i}}^{*}
\sigma_{b_{\f}d_{\i}}\sigma_{a_{\f}d_{\i}}^{*}|=B^{-2}$ which means that the
two sums in the second line of Eq.~(\ref{step2}) yield only a negligible
correction ${\cal O}(B^{-1})$ because the number of terms are $N^3(N-1)$ and
$N(N-1)(B-N)$, respectively.  It is the first line in Eq.~(\ref{step2}) which
gives the dominant contribution. With exactly the same methods we can consider
the contribution from special cases of the diagram in Fig.~\ref{arrows}d where
the length of one of the subsequences $a,b,c,d$ vanishes. We find that for
vanishing $a$ or $b$ we get $t^{(4)}_{bcd}=t^{(4)}_{acd}=-t^{(4)}_{abcd}$
while for vanishing $c$ or $d$ no contribution results. Finally we have
$\<\tr\,(tt^{\dagger})^{2}\>=t^{(4)}_{abcd}+t^{(4)}_{bcd}+t^{(4)}_{acd}=
-{N_{1}^{2}N_{2}^{2}/
  N^3}$ and substitution of this result into Eq.~(\ref{P}) shows that we have
indeed reproduced the RMT result from correlated classical trajectories.

We have checked that our result remains unchanged if we
substitute in Eq.~(\ref{P})
$\tr\,tt^{\dagger}(1-tt^{\dagger})=\tr\,tt^{\dagger}rr^{\dagger}$. Quantum
mechanically this is just a consequence of unitarity, but within the
semiclassical approach it is a nontrivial result since entirely different
trajectories contribute and unitarity is restored only if all relevant
correlations between them are
properly accounted for. 

\acknowledgments We are grateful to O.~Agam and H.~Schomerus for stimulating
our interest in the problem.

\end{document}